\documentclass[fleqn, usenatbib, twocolumns]{mnras}

\usepackage{newtxtext,newtxmath}
\usepackage[T1]{fontenc}
\usepackage{soul}

\DeclareRobustCommand{\VAN}[3]{#2}
\let\VANthebibliography\thebibliography
\def\thebibliography{\DeclareRobustCommand{\VAN}[3]{##3}\VANthebibliography}

\usepackage{graphicx}
\usepackage{amsmath}	

\newcommand{\angstrom}{\text{\normalfont\AA}}

\newcommand{\wdbd}[1]{VVV J1438-6158 AB} 
\newcommand{\wdbdA}[1]{VVV J1438-6158 A} 
\newcommand{\wdbdB}[1]{VVV J1438-6158 B} 

\title[\wdbd{}]{A Benchmark White Dwarf--Ultracool Dwarf Wide Field Binary}

\author[Ferreira et al.]{Thiago Ferreira,$^{1, 2}$\thanks{E-mail: \href{mailto:thiago.dossantos@yale.edu}{thiago.dossantos@yale.edu}}, Roberto K. Saito$^{3}$, Dante Minniti$^{4,5,3}$, Andrea Mejías$^{4}$, Claudio Caceres$^{4, 6}$, \newauthor Javier Alonso-García$^{7, 8}$, Juan Carlos Beamín$^{4}$, Leigh C. Smith$^{9}$, Matías Gomez$^{4}$, \newauthor Philip W. Lucas$^{10}$, and Valentin D. Ivanov$^{11}$\\
\\
${^1}$ Department of Astronomy, Yale University, 219 Prospect St., New Haven, CT 06511, USA. \\
$^{2}$ Instituto de Astronomia, Geofísica e Ciências Atmosféricas, Universidade de São Paulo, 05508-090, São Paulo, Brazil. \\
$^{3}$ Departamento de Física, Universidade Federal de Santa Catarina, Trindade 88040-900, Florianópolis, Brazil. \\
$^{4}$ Instituto de Astrof\'isica, Facultad de Ciencias Exactas, Universidad Andres Bello, Av. Fern\'andez Concha 700, Santiago, Chile. \\
$^{5}$ Vatican Observatory, V00120 Vatican City State, Italy. \\
$^{6}$ Núcleo Milenio de Formación Planetaria – NPF, Valparaíso, Chile \\
$^{7}$ Centro de Astronomía (CITEVA), Universidad de Antofagasta, Av. Angamos 601, Antofagasta, Chile. \\
$^{8}$ Millennium Institute of Astrophysics, Nuncio Monseñor Sotero Sanz 100, Of. 104, Providencia, Santiago, Chile. \\
$^{9}$ Institute of Astronomy, University of Cambridge, Madingley Road, Cambridge CB3 0HA, UK. \\
$^{10}$ Centre for Astrophysics, University of Hertfordshire, College Lane, Hatfield AL10 9AB, UK. \\
$^{11}$ European Southern Observatory, Karl-Schwarzschild-Stra{\ss}e 2, D-85748 Garching bei München, Germany.}

\date{Accepted XXX. Received YYY; in original form ZZZ}

\pubyear{2023}

\begin{document}

\label{firstpage} \pagerange{\pageref{firstpage}--\pageref{lastpage}} \maketitle

\begin{abstract} 

{We present the discovery and multi-wavelength characterisation of \wdbd{}, a new field wide-binary system consisting of a $4.6^{+5.5}_{-2.4}~{\rm Gyr}$ and $T_{\rm eff} = 9500\pm125~K$ DA white dwarf (WD) and a $T_{\rm eff} = 2400\pm50~K$ M8 ultracool dwarf (UCD). The projected separation of the system is $a = 1236.73~{\rm au}$ ($\sim13.8"$), and although along the line-of-sight towards the Scorpius-Centaurus (Sco-Cen) stellar association, \wdbd{} is likely to be a field star, from a kinematic 6D probabilistic analysis. We estimated the physical, and dynamical parameters of both components via interpolations with theoretical models and evolutionary tracks, which allowed us to retrieve a mass of $0.62\pm0.18~M_\odot$ for the WD, and a mass of $98.5\pm6.2~M_{\rm Jup}$ ($\sim0.094\pm0.006~M_\odot$) for the UCD. The radii of the two components were also estimated at $0.01309\pm0.0003~R_{\odot}$ and $1.22\pm0.05~R_{\rm Jup}$, respectively. \wdbd{} stands out as a benchmark system for comprehending the evolution of WDs and low-mass companions given its status as one of the most widely separated WD+UCD systems known to date, which likely indicates that both components may have evolved independently of each other, and also being characterised by a large mass-ratio ($q = 0.15\pm0.04$), which likely indicates a formation pathway similar to that of stellar binary systems.}

\end{abstract}


\section{Introduction}

{Binary systems involving a white dwarf (WD) and a companion of extremely low-mass, such as an ultracool dwarf (UCD) or brown dwarf (BD), present a unique opportunity to investigate the evolution of low-mass stars and sub-stellar objects. The working definition and distinction between BDs ($T_{\rm eff}~\leq~2200~K$), and UCDs ($T_{\rm eff}~\leq~2700~K$) are based on their masses and temperatures. Both surpass the planetary threshold for deuterium burning in their cores ($\sim~14-16~M_{\rm Jup}$; \citealt{2011ApJ...727...57S}, but BDs fall short of initiating hydrogen nuclear fusion ($\sim80-85~M_{\rm Jup}$; \citealt{1963ApJ...137.1121K, 2001RvMP...73..719B}). In the upper span of their mass spectrum ($\gtrapprox~60~M_{\rm Jup}$), their sizes are predominantly regulated by electron-degeneracy pressure, alike WDs, while at the lower limit of this range ($\sim~14-20~M_{\rm Jup}$), their sizes are mainly determined by Coulomb pressure \citep{2006AREPS..34..193B}. WDs, on the other hand, are remnants of stars with masses in the Main Sequence ranging from $0.8$ to $7.5$ $M_\odot$ \citep{1968ARA&A...6..351W, 2022PhR...988....1S}, composed primarily of degenerate matter, and are supported against gravitational collapse by electron-degeneracy pressure.}

Significant progress has been achieved in the investigation of cataclysmic variable stars (CVs) involving stellar remnants and sub-stellar objects (e.g., \citealt{2019MNRAS.484.2566L, 2022MNRAS.513.3587Z}, and references therein), however, many of these systems are distinguished by their close proximity, often leading to common envelope interactions and profound influence on each other component's evolution.

\wdbd{} was discovered from the investigation of low-mass stars and free-floating planets (FFPs) towards the young Lower Centaurus Crux stellar association (LCC; \citealt{1999AJ....117..354D}) by \cite{2022cosp...44..591M}. Differing from previously known WD+UCD/BD binaries (e.g., \citealt{2004AJ....128.1868F, 2005ApJS..161..394F, 2005ApJ...630L.173S, 2009A&A...500.1207S, 2012ApJ...759L..34C, 2013MNRAS.429.3492S, 2013A&A...558A..96B, 2013MmSAI..84.1027C, 2014MNRAS.445.2106L, 2019MNRAS.484.2566L, 2020MNRAS.497.3571C}), and also from the first BD discovered by the VVV survey not belonging to a binary system (VVV BD001; \citealt{2013A&A...557L...8B}), in terms of orbital separation, \wdbd{} resides at the opposite end of the spectrum, exhibiting a notably loose configuration wherein both components appear to have evolved independently from one another, i.e., without exerting any discernible influence on each other's evolutionary paths. Its significance relies upon the scarcity, albeit non-absence, of systems with well-established ages similar to it. A similar WD+UCD binary system, WD0837+185, was also discovered within the Praesepe cluster \citep{2012ApJ...759L..34C}, however, it differs from the system reported here with its much closer projected separation of $\sim0.006~{\rm au}$ and age of $625$ Myr, which likely implies interaction, making unlikely the independent evolution of its components. {As of the current writing, there are 8 known WD+UCD/BD binaries, spanning a range from relative proximity between the components ($a < 100~{\rm au}$; \citealt{2013MNRAS.429.3492S}), or exhibiting an intermediate configuration (e.g., \citealt{1988Natur.336..656B, 2020ApJ...899..123M, 2023MNRAS.519.5008F}), to loose separations ($a > 1000~{\rm au}$; e.g., \citealt{2011ApJ...732L..29R, 2011ApJ...730L...9L, 2011MNRAS.410..705D, 2014ApJ...792..119D}).}

In this paper, we present the characterisation of \wdbd{}, a new WD+UCD binary system towards the young Scorpius-Centaurus (Sco-Cen) stellar association. In Section \ref{sec:obs}, we describe the observational data used in the analysis, focusing on the multi-colour photometry obtained for both components: $G$, $BP$, and $RP$ magnitudes from \emph{Gaia} DR3 \citep{2021A&A...649A...1G}, $grizY$ magnitudes from DECam Plane Survey (DECaPS; \citealt{2018ApJS..234...39S}), and $ZYJHK_\mathrm{s}$ magnitudes from the VISTA Variables in the Via Lactea survey (VVV; \citealt{2010NewA...15..433M}). In Section \ref{sec:physical}, we present the physical parameters of WD and its companion, including effective temperatures, masses, radii, surface gravities, and ages from interpolation with theoretical models. In Section \ref{sec:membership}, we briefly discuss the likelihood of membership to Sco-Cen by \wdbd{}. A light curve analysis of the system in the VVV-$K_\mathrm{s}$ pass-band is discussed in Section \ref{sec:lcs}, and in Section \ref{sec:conclusions}, we present our conclusions, summarising the key findings from our analysis. 

\section{Observations of \wdbd{}}\label{sec:obs}

{\wdbd{} was identified by \cite{2021MNRAS.506.2269E} as a co-moving pair with a small chance alignment probability $\mathcal{R} = 3.015\times10^{-5}$, who classified it as a co-moving pair composed of a WD and a main-sequence (MS) star. Independently, we classified the system as a field binary consisting of a WD and a UCD towards the Sco-Cen stellar association from data of the VVV survey and \emph{Gaia} DR3. In the absence of radial velocity measurements for both components of \wdbd{}, we cannot estimate the 3D space velocity, nor the difference, for the system. Nevertheless, the high-bound probability assumption for the system relies on $\mathcal{R}~\ll~0.1$, which corresponds approximately to $\geq$90\% bound probability (e.g., \citealt{2021MNRAS.506.2269E})}. Table \ref{tab:gaia_WDBD} presents the \emph{Gaia} DR3 measurements for both components, encompassing locations, parallaxes, proper motions in equatorial components, the renormalised unit weight errors (RUWEs), and distances derived from the parallax measurements. 

We adopt a distance of $89.6\pm0.5$ pc for this system based on the more accurately measured parallax value of the WD component, which has an uncertainty of $0.07$ mas. Its Galactic space velocities components are $U = -0.508~km~s^{-1}$, $V = 4.392~km~s^{-1}$ and $W = -1.569~km~s^{-1}$ from \cite{2019MNRAS.485.5573T}, which suggests that the system belongs to the thin disc. {The tangential velocity of both components, $V_{\rm t} = 4.74\cdot\mu_{\rm total}\cdot\Delta$, was estimated at $V_{\rm t, WD} = 14.62\pm0.02$ $km~s^{-1}$ and $V_{\rm t, UCD} = 17.87\pm0.15$ $km~s^{-1}$}. The projected separation between the WD and UCD is $13.8"$, which corresponds to a physical separation of $\sim1240$ au at that distance, and note that the RUWE values for both components are relatively low, which indicates reliable astrometric estimations from \emph{Gaia} \citep{2021A&A...649A...2L}, {and that both components do not likely possess close stellar companions \citep{2022RNAAS...6...18F}}. 

\begin{table*}
	\centering
	\caption{\emph{Gaia} DR3 astrometric measurements and derived parameters for \wdbd{}. \\ 
             {\it Note}: Component A (first row) corresponds to the WD, while component B (second row) corresponds to the UCD.}
	\label{tab:gaia_WDBD}
	\begin{tabular}{lcccccccc}
		\hline
		\emph{Gaia} DR3 ID & $\alpha$ & $\delta$ & $\overline\pi$ & $\mu_{\alpha}\cos\delta$ & $\mu_{\delta}$ & RUWE & $\Delta$ \\
          & (deg) & (deg) & (mas) & (mas yr$^{-1}$) & (mas yr$^{-1}$) & & (pc)\\
		\hline
        $5877522801734482432$ & $219.624\pm0.0387$ & $-61.9806\pm0.0446$ & $11.162\pm0.066$ & $-14.975\pm0.062$ & $-31.019\pm0.060$ & $0.956$ & $89.6\pm0.5$ \\[0.1cm]
        $5877522797372878208$ & $219.628\pm0.2409$ & $-61.9840\pm0.2704$ & $11.545\pm0.389$ & $-14.231\pm0.378$ & $-29.604\pm0.372$ & $1.180$ & $86.6\pm2.9$ \\[0.1cm]
        \hline
	\end{tabular}
\end{table*}

\begin{figure*}
    \centering
    \includegraphics[width = 0.197\linewidth]{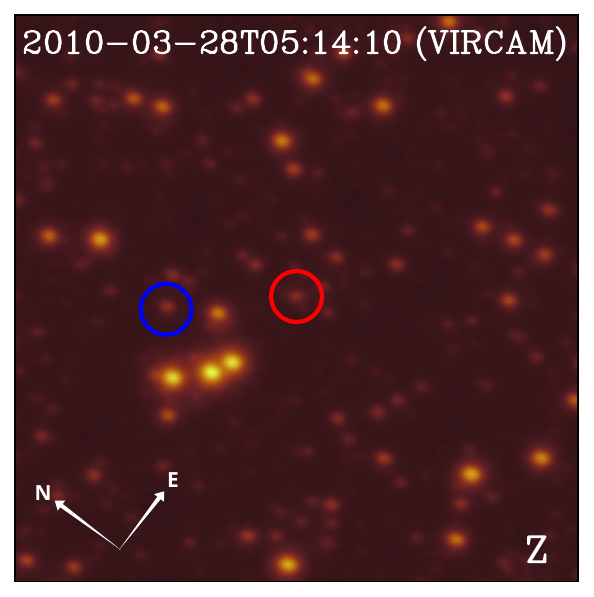}
    \includegraphics[width = 0.197\linewidth]{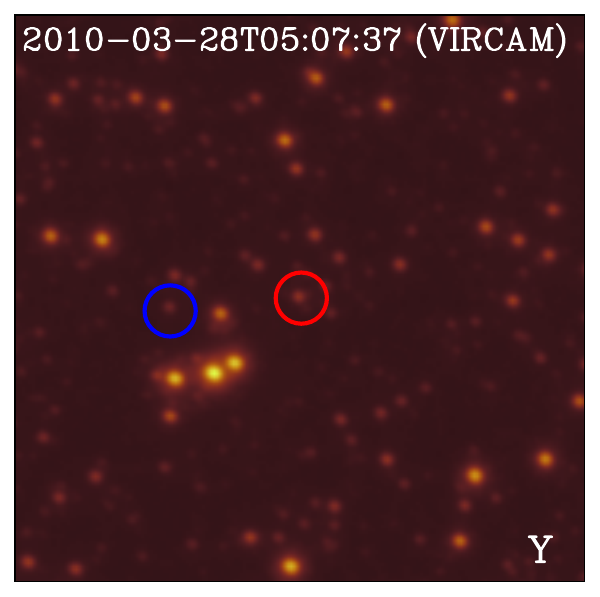}
    \includegraphics[width = 0.197\linewidth]{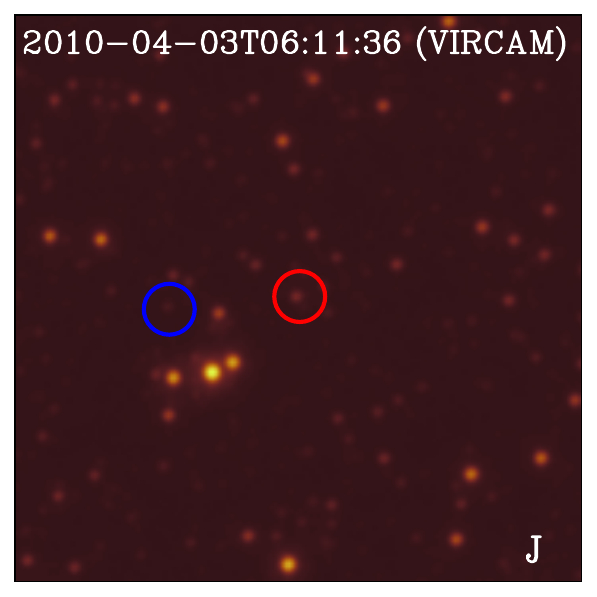}
    \includegraphics[width = 0.197\linewidth]{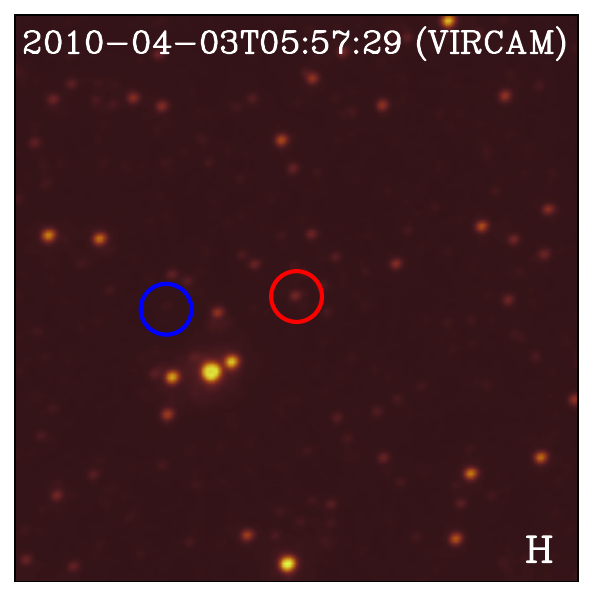}
    \includegraphics[width = 0.197\linewidth]{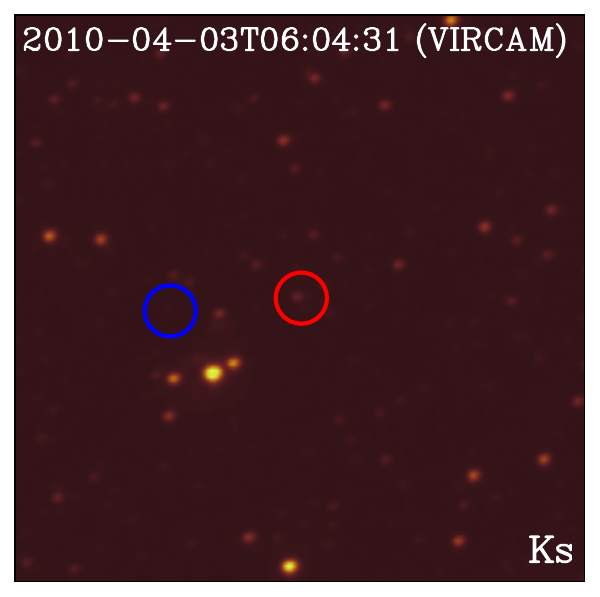}
    \caption{Near-IR finding charts from the VVV DR5 of the WD (component A; blue circles), and its UCD companion (component B; red circles) in $ZYJHK_\mathrm{s}$ pass-bands in 2010. Despite the crowded nature of the field, the binary components are well-separated ($13.8"$), and visibly, the UCD is brighter than the WD at longer near-IR wavelengths. {All images have the same orientation as indicated in the first panel.}}
    \label{fig:VISTAstamps}
\end{figure*}

Although the WD companion was too faint to be detected during observations of the 2MASS survey \citep{2006AJ....131.1163S}, the UCD was detected (source ID 2MASS J14383082-6159022). For both components, however, near-infrared (near-IR) $ZYJHK_\mathrm{s}$ photometry was obtained by the VVV survey \citep{2010NewA...15..433M}. They are situated within the overlapping region of the Galactic disc VVV tiles d014 and d015, and despite the crowded nature of the field, the optical and near-IR images suggest that they remain relatively uncontaminated (see Figure \ref{fig:VISTAstamps}). 

\begin{figure*}
    \centering
    \includegraphics[width = 0.33\linewidth]{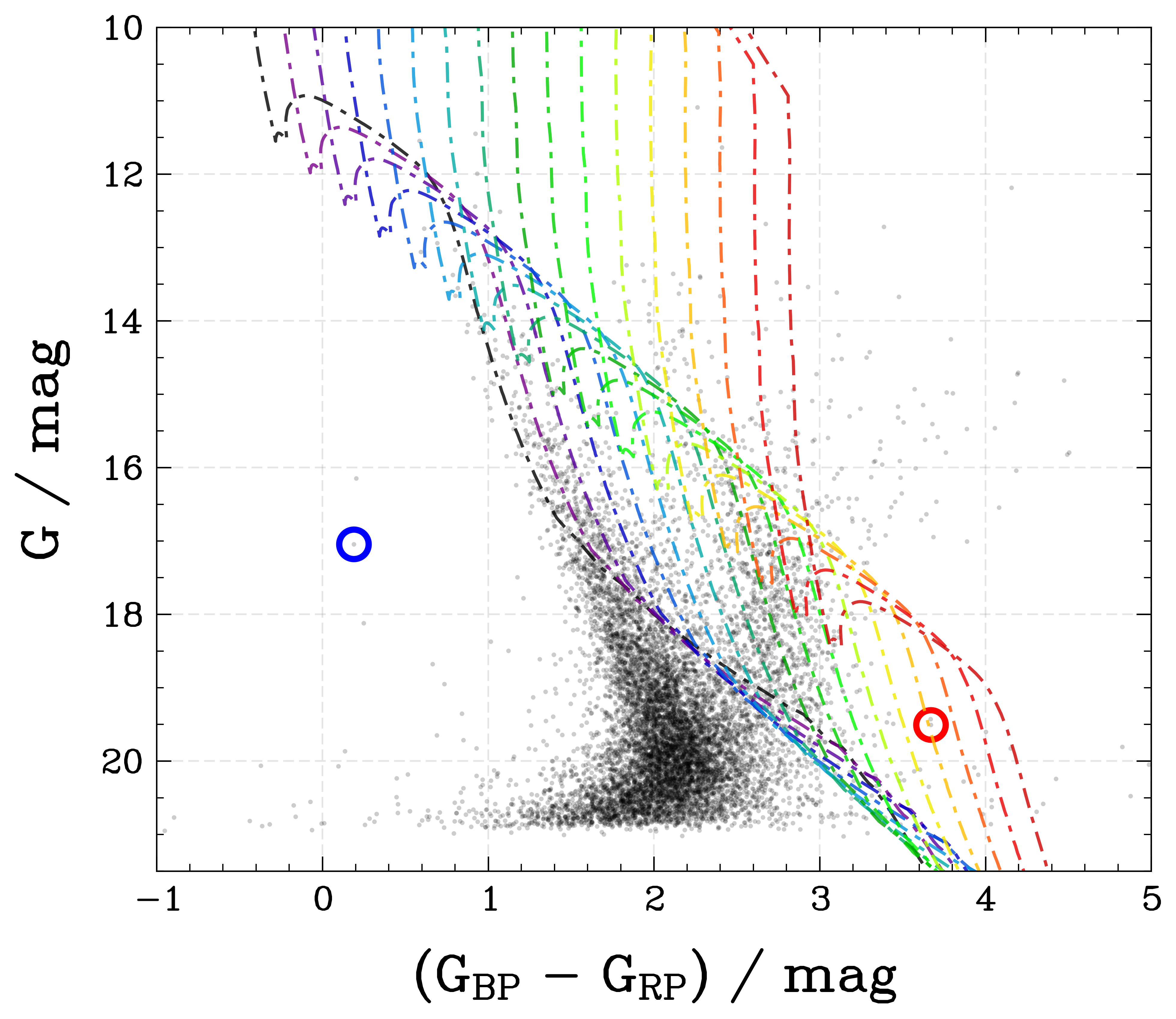}
    \includegraphics[width = 0.33\linewidth]{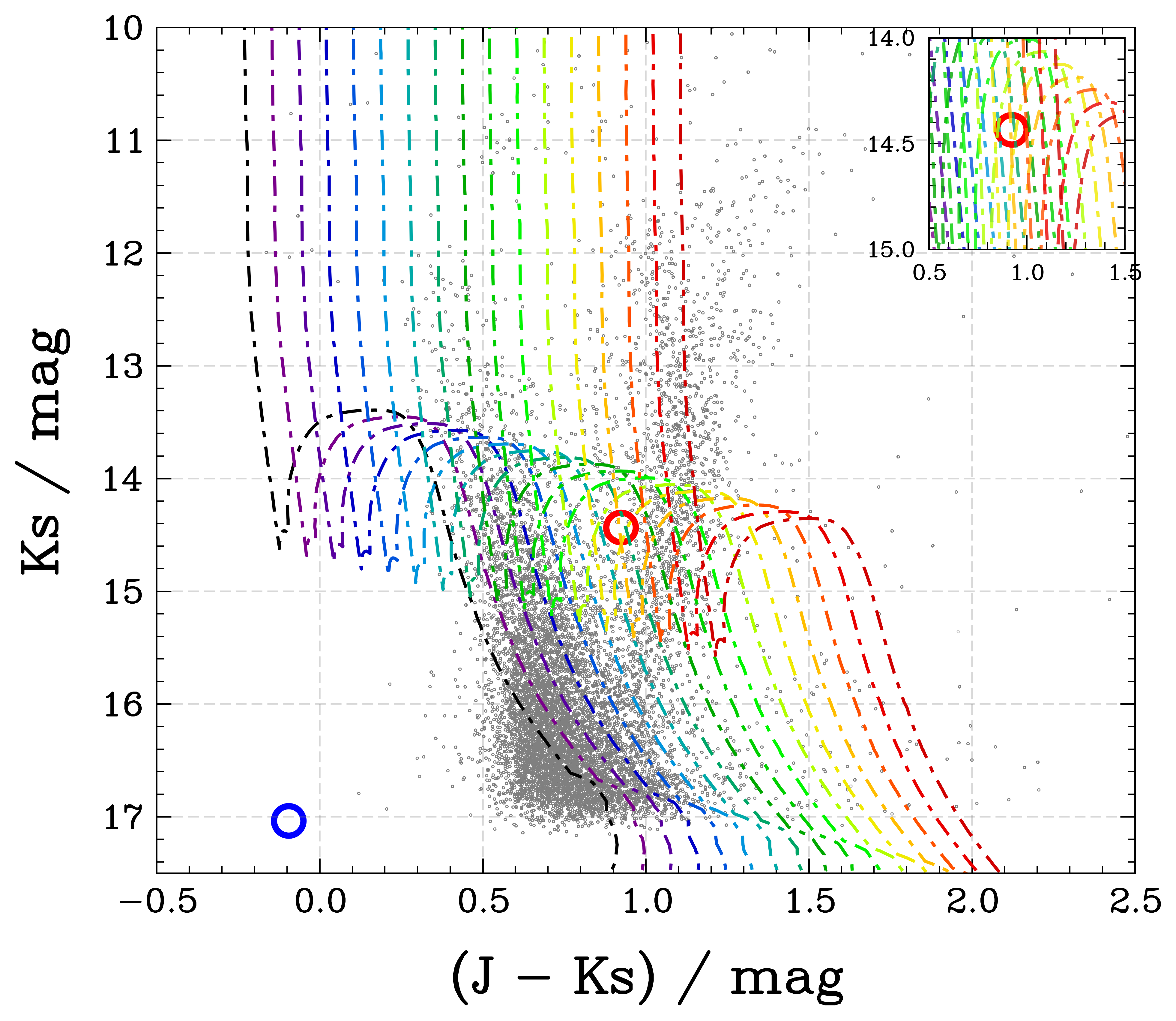}
    \includegraphics[width = 0.33\linewidth]{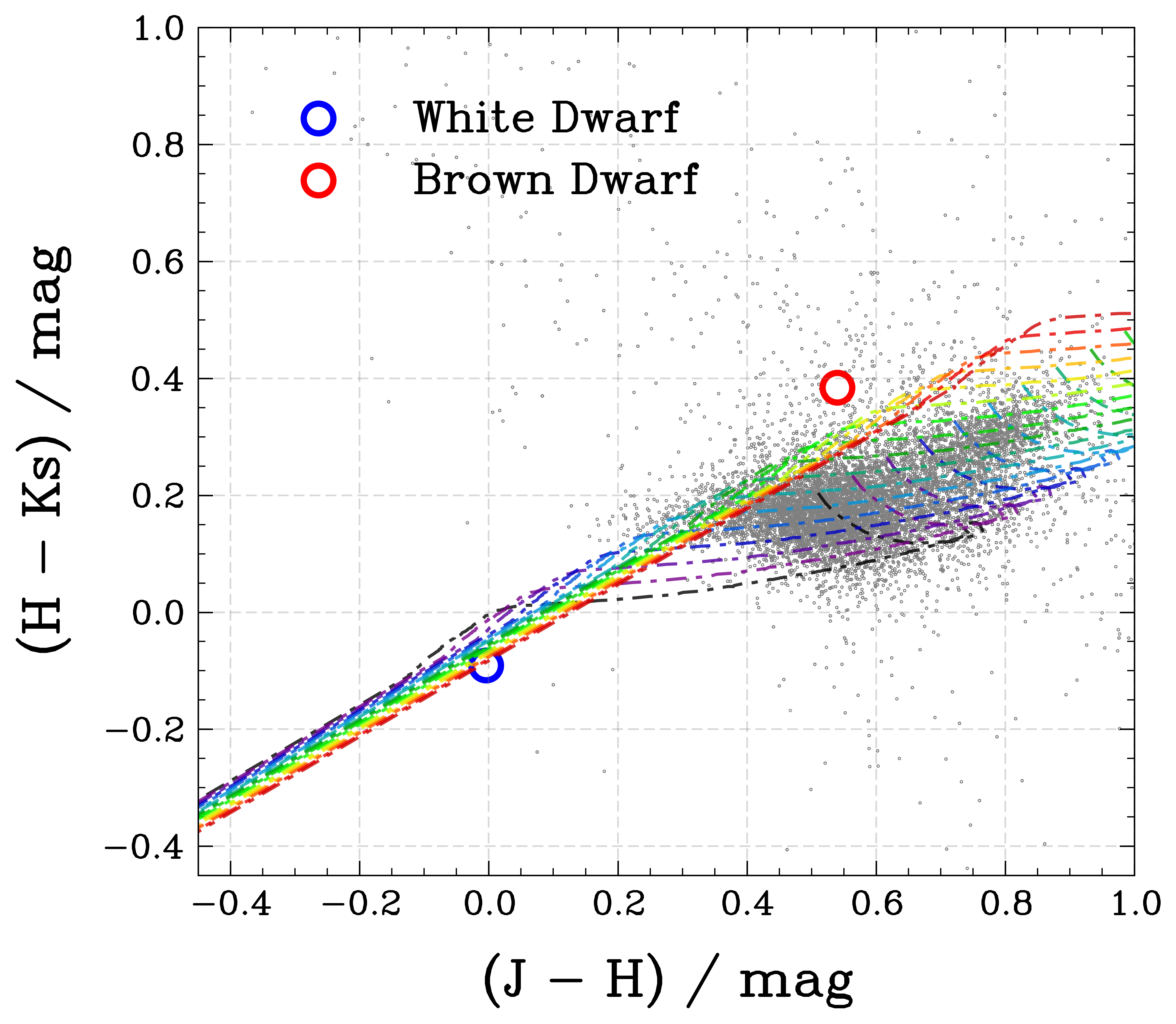}
    \caption{Optical \emph{Gaia} $G$ vs. $BP-RP$ (first panel), near-IR VVV $K_\mathrm{s}$ vs. $J - K_\mathrm{s}$ CMD (second panel), and $J - H$ vs. $H - K_\mathrm{s}$ CCD (third panel) diagrams for a $5$ arcmin field centred on the WD+UCD binary pair. The WD is shown with a blue circle and the UCD with a red circle. Evolutionary tracks for distinct $A_V$ in each panel are from PARSEC v1.2S + COLIBRI S$_{35}$.}
    \label{fig:CMDs}
\end{figure*}

Figure \ref{fig:CMDs} depicts the optical and near-IR colour-magnitude (CMD) and colour-colour diagrams (CCD) of the surrounding field ($5$ arcmin of the central location of the WD) containing the targets. In both diagrams, the two components are separated. Solar metallicity isochrones\footnote{\href{http://stev.oapd.inaf.it/cgi-bin/cmd}{\tt http://stev.oapd.inaf.it/cgi-bin/cmd}} (PARSEC v1.2S and COLIBRI S$_{35}$ tracks; \citealt{2014MNRAS.444.2525C, 2019MNRAS.485.5666P}) for ages of $1-10$ Gyr, extinctions $A_V$ ranging from $0-8$ mag, shifted to the distance of the Galactic centre, $\approx8.7$ kpc \citep{2009A&A...498...95V}, are overlaid to the CMDs and CCD. \wdbd{} is located in a region of relatively high reddening\footnote{\href{https://irsa.ipac.caltech.edu/applications/DUST/}{\tt https://irsa.ipac.caltech.edu/DUST/}}, with a measured colour excess E(B$-$V)$_{\rm SFD} = 1.75\pm0.06$ mag, and visual extinction $A_{\rm V, SFD} = 5.42\pm0.19$ mag \citep{1998ApJ...500..525S, 2011ApJ...737..103S}, while the colour excess at the distance of the system (E(B$-$V)$_{\Delta}$) is $0.012\pm0.017$ mag and the visual extinction also at the distance of the system ($A_{\rm V, \Delta}$) is $0.037\pm0.052$ mag \citep{2017A&A...606A..65C}\footnote{\href{https://stilism.obspm.fr/}{\tt https://stilism.obspm.fr/}}. 

Table \ref{tab:photometry} presents the photometric data for the WD and UCD obtained from various optical and near-IR catalogues, including the \emph{Gaia} DR3, DECaPS, VVV, and 2MASS \citep{2006AJ....131.1163S}. The absolute magnitudes in \emph{Gaia}$-G$, and VVVX$-J/K_\mathrm{s}$ bands, assuming a distance of $\Delta = 89.6$ pc for this system, are provided in the last rows. Conservative errors were considered, primarily influenced by the measured parallax difference between both components. It is worth noting the absence of near-IR excess in the WD's colours, which suggests the absence of an additional close companion, or a circumstellar disc. Moreover, a thorough examination of the near-IR CCD diagram presented in \cite{2011MNRAS.416.2768S} (refer to their Figure 1, which also relies on the findings of \citealt{2005PASP..117.1378W}), confirms that the WD occupies a distinct position within the solitary WD region. \\

\begin{table}
	\centering
	\caption{Optical and near-IR photometry of \wdbd{} binary.}
	\label{tab:photometry}
	\begin{tabular}{llcc}
		\hline
        Survey & Filter & WD (A) & UCD (B) \\
		\hline
        \emph{Gaia} DR3 (Vega) & $G$ & $17.05\pm0.01$ & $19.51\pm0.01$\\[0.1cm]
        \emph{Gaia} DR3 (Vega) & $BP$ & $17.12\pm0.01$ & $21.43\pm0.22$\\[0.1cm]
        \emph{Gaia} DR3 (Vega) & $RP$ & $16.93\pm0.01$ & $17.76\pm0.01$\\[0.1cm]
        VVV (Vega) & $J$ & $16.94\pm0.03$ & $15.43\pm0.02$\\[0.1cm]
        VVV (Vega) & $H$ & $16.87\pm0.03$ & $14.87\pm0.02$\\[0.1cm]
        VVV (Vega) & $K_\mathrm{s}$ & $16.89\pm0.03$ & $14.55\pm0.02$\\[0.1cm]
        2MASS (Vega) & $J$ & $-$ & $15.38\pm0.02$\\[0.1cm]
        2MASS (Vega) & $H$ & $-$ & $14.74\pm0.03$\\[0.1cm]
        2MASS (Vega) & $K_\mathrm{s}$ & $-$ & $13.93\pm0.05$\\[0.1cm]
        DECaPS (AB) & $g$ & $17.05\pm0.01$ & $23.34\pm0.10$\\[0.1cm]
        DECaPS (AB) & $r$ & $17.13\pm0.01$ & $21.47\pm0.05$\\[0.1cm]
        DECaPS (AB) & $i$ & $17.24\pm0.01$ & $18.43\pm0.01$\\[0.1cm]
        DECaPS (AB) & $z$ & $17.40\pm0.01$ & $17.25\pm0.01$\\[0.1cm]
        DECaPS (AB) & $Y$ & $17.49\pm0.01$ & $16.88\pm0.01$\\[0.1cm]
        \hline
        \emph{Gaia} DR3 & $M_{\tiny G}$ & $12.28\pm0.07$ & $14.76\pm0.07$ \\[0.1cm]
        VVVX (Vega) & $M_{\tiny J}$ & $12.17\pm0.08$ & $10.65\pm0.07$ \\[0.1cm]
        VVVX (Vega) & $M_{\tiny K_\mathrm{s}}$ & $12.47\pm0.08$ & $13.66\pm0.07$ \\[0.1cm]
        \hline 
	\end{tabular}
\end{table}

\section{Spectral Energy Distribution (SED) Analysis} \label{sec:physical}

\begin{figure}
    \centering
    \includegraphics[width = \columnwidth]{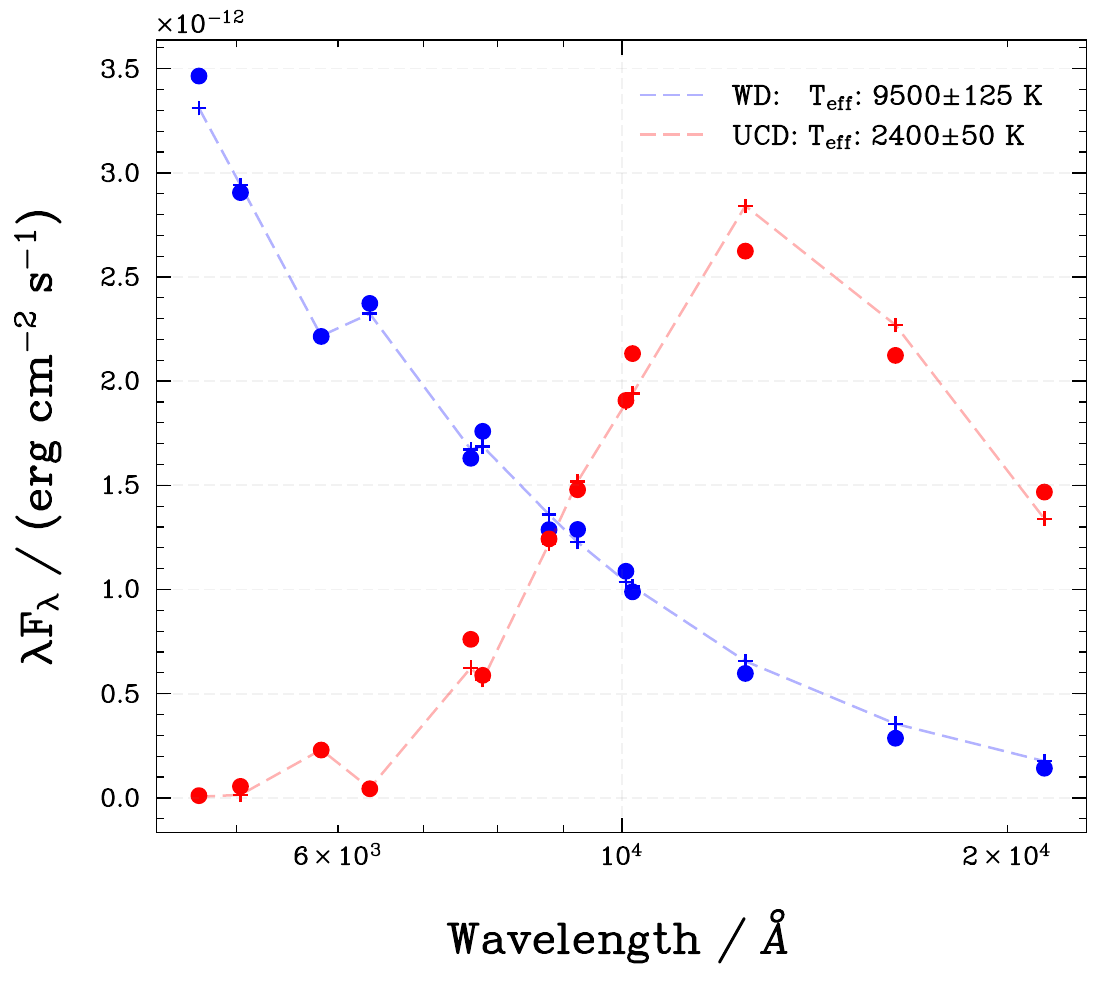}
    \caption{SED fitting of \wdbd{} pair. The circles correspond to the observed flux for both components (red for the UCD and blue for the WD), while the dashed lines plus crosses correspond to the respective best-fit models.}
    \label{fig:SED-VOSA}
\end{figure}

We integrated the observed fluxes from the optical ($\lambda_{\rm min} = 4673.24~\angstrom$) to IR-region ($\lambda_{\rm max} = 21376~\angstrom$) using the Virtual Observatory Spectral Energy Distribution (SED) Analyser\footnote{\href{http://svo2.cab.inta-csic.es/theory/vosa/}{\tt http://svo2.cab.inta-csic.es/theory/vosa/}} web tool (VOSA; \citealt{2008A&A...492..277B}) to estimate the physical and atmospheric parameters of both components of \wdbd{} system. Figure \ref{fig:SED-VOSA} depicts their multi-frequency SED collected from the \emph{Gaia}, DECaPS, and VVV surveys, and Table \ref{tab:wdbdphysics} summarises the main properties of both components of \wdbd{} binary system. 

\begin{table}
	\centering
    \caption{Fundamental properties of \wdbd{} binary retrieved with SED analysis.}
	\label{tab:wdbdphysics}
	\begin{tabular}{l|lr}
		\hline
         & Value & Description \\
		\hline
        {\it White Dwarf} & & \\[0.1cm]
        Sp. T. & DA & Spectral type \\[0.1cm]
        $T_{\rm eff}$ (K) & $9500\pm125$ & Effective temperature \\[0.1cm]
        $\log(g)$ (cgs) & $8.00\pm0.12$ & Surface gravity \\[0.1cm]
        $L_{\rm bol}$ ($L_\odot$) & $(1.257\pm0.005)\times10^{-3}$ & Bolometric luminosity \\[0.1cm]
        $M$ ($M_\odot$) & $0.62\pm0.18$ & Mass (Solar units) \\[0.1cm]
        $R$ ($R_\odot$) & $0.01309\pm0.0003$ & Radius (Solar units) \\[0.1cm]
        $R$ ($R_\oplus$) & $1.42\pm0.03$ & Radius (Earth units) \\[0.1cm]
        $\tau$ (Gyr) & $0.74^{+0.09}_{-0.06}$ & Cooling age \\[0.1cm]
        $\Gamma$ (Gyr) & $4.6^{+5.5}_{-2.4}$ & WD's total age \\[0.1cm]
        $\Gamma'$ (Gyr) & $3.8^{+5.6}_{-2.5}$ & Main Sequence age \\[0.1cm]
        $M_{\rm init}$ ($M_\odot$) & $1.37^{+0.63}_{-0.32}$ & WD Progenitor's mass \\[0.1cm]
        \hline 
        {\it Ultracool Dwarf} & & \\[0.1cm]
        Sp. Type & M8 & Spectral type \\[0.1cm]
        $T_{\rm eff}$ (K) & $2400\pm50$ & Effective temperature \\[0.1cm]
        $\log(g)$ (cgs) & $5.21\pm0.04$ & Surface gravity \\[0.1cm]
        $L_{\rm bol}$ ($L_\odot$) & $(5.60\pm0.02)\times10^{-4}$ & Bolometric luminosity \\[0.1cm]
        $M$ ($M_\odot$) & $0.094\pm0.006$ & Mass (Solar units) \\[0.1cm]
        $M$ ($M_{\rm Jup}$) & $98.47\pm6.28$ & Mass (Jupiter units) \\[0.1cm]
        $R$ ($R_{\rm Jup}$) & $1.22\pm0.05$ & Radius (Jupiter units) \\[0.1cm]
        $R$ ($R_\odot$) & $0.1253\pm0.005$ & Radius (Solar units) \\[0.1cm]
        $\rho$ (g cm$^{-3}$) & $67.25\pm9.31$ & Density \\[0.1cm]
        \hline 
        $P_{\rm orb}$ (yrs) & $(141.85\pm4.52)\times10^{3}$ & Orbital period (years) \\[0.1cm]
        $\Omega$ (au) & $1236.73$ & Projected separation \\[0.1cm]
        $\mathcal{R}$ & $3.015\times10^{-5}$ & Chance alignment prob. \\[0.02cm]
        \hline
    \end{tabular}
\end{table} 

\subsection{White Dwarf's characterisation}

In the \emph{Gaia} EDR3 catalogue, \wdbdA{} is classified as a WD with 99.47\% probability \citep{2021MNRAS.508.3877G}, {and have 79.3\% probability of being a DA-type WD (with a predominantly hydrogen-rich atmosphere; \citealt{2023MNRAS.518.5106J})}. Pure-hydrogen and pure helium atmospheric models from \cite{2021MNRAS.508.3877G} yielded effective temperatures of $T_{\rm eff, H/He} = 9588.61\pm123.51/9405.41\pm128.14~K$, surface gravities at $\log(g)_{\rm H/He} = 7.98\pm0.04/7.89\pm0.04$, and masses $M_{\rm H/He} = 0.58\pm0.02/0.51\pm0.02~M_\odot$. 

The best fit to the available photometric measurements of \wdbdA{} (Table \ref{tab:photometry}) was obtained with the WD spectra and atmosphere model library presented in \cite{2010MmSAI..81..921K}, yielding an effective temperature of $T_{\rm eff} = 9500\pm125~K$ and surface gravity $\log(g) = 8.00\pm0.12$, with $\chi_\nu^2 = 1.38$ (11 free parameters). The bolometric luminosity for the adopted distance is $L_{\rm bol} = (1.257\pm0.005)\times10^{-3}~L_\odot$, and applying the Stefan-Boltzmann law led to an estimation of the object's radius as $R = 0.01309\pm0.0003~R_\odot$. From the best-fit surface gravity and estimated radius, the WD's mass was determined as $M = 0.63\pm0.18~M_\odot$. Within their uncertainties, these WD parameters appear to be similar to what has been derived for similar objects (e.g., \citealt{1935MNRAS..95..207C, 1958ses..book.....S, 1969stph.book.....L}). Alternatively, a direct black body fit resulted in $T_{\rm eff, BB} = 9900\pm25~K$, and $L_{\rm bol, BB} = 1.257\times10^{-3}\pm1.6\times10^{-6}~L_\odot$.

\subsubsection{WD's Age and Progenitor mass}

To estimate the age of \wdbdA{}, we used the {\sc wdwarfdate} software \citep{2022AJ....164...62K}, which adopts a Bayesian approach for thick hydrogen (DA) and non-DA layer, taking into account the remnant's effective temperature and surface gravity. This approach allowed us to estimate the final mass of the WD and the initial mass of its progenitor. For a DA WD, the main-sequence age and mass were derived as $\Gamma_{\rm DA} = 3.88^{+5.57}_{-2.51}$ Gyr and $M_{\rm init, DA} = 1.37^{+0.63}_{-0.32}~M_\odot$, respectively, from an initial-final mass relation model of \cite{2018ApJ...866...21C}, a cooling model from \cite{2020ApJ...901...93B}, and evolutionary tracks from \cite{2016ApJ...823..102C} and \cite{2016ApJS..222....8D}. 

\begin{figure}
    \centering
    \includegraphics[width = \columnwidth]{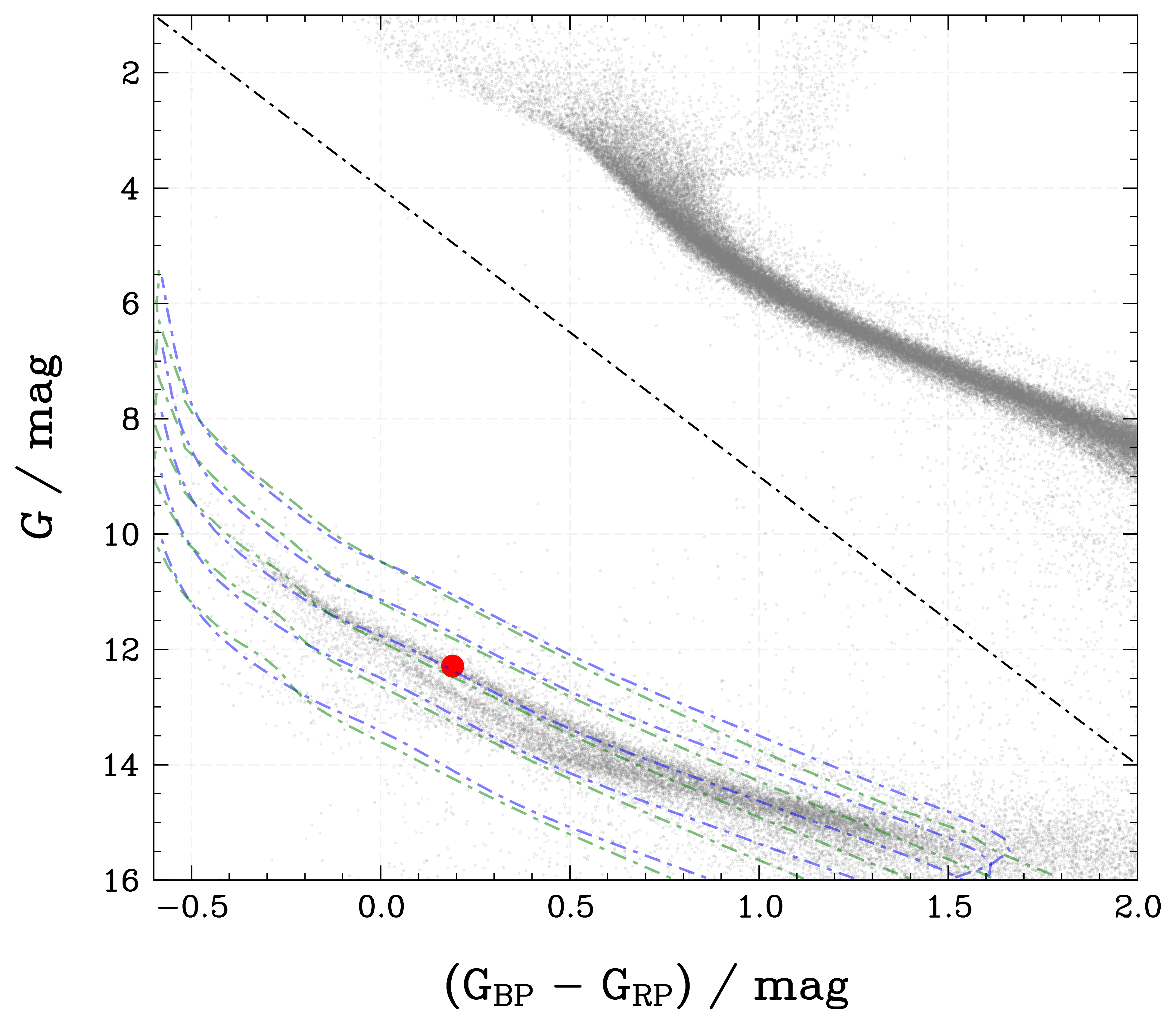}
    \caption{Optical \emph{Gaia} absolute $G$ vs. $BP-RP$ CMD diagram for approximately 370.000 nearby ($\overline\pi < 100^{-1}$ mas) sources with high-significance parallax ($\overline\pi/\sigma_{\overline\pi} > 100$; grey-coloured background). \wdbdA{} is indicated as the red dot. The blue and green dashed lines indicate evolutionary cooling tracks for DA-type (blue) and DB-type (green) WD stars with surface gravity ranging from $\log(g) = 7.0$ (top) to $\log(g) = 9.0$ (bottom; $\delta \log(g) = 0.5$).}
    \label{fig:CMD_CT}
\end{figure}

Figure \ref{fig:CMD_CT} depicts the optical $G$ vs. $G_{\rm BP} - G_{\rm RP}$ CMD for nearby objects presented in the \emph{Gaia} catalogue, and the positioning of \wdbdA{} in such a diagram. Overlaid to the CMD are indicated evolutionary cooling tracks\footnote{\href{https://www.astro.umontreal.ca/~bergeron/CoolingModels/}{\tt https://www.astro.umontreal.ca/$\sim$bergeron/CoolingModels/}} for pure-hydrogen DA- and pure-helium DB-type WDs with $\log(g) = 7.5-9.0$ from \cite{2006AJ....132.1221H} and \cite{2020ApJ...901...93B}. 

With {\sc wdwarfdate} we obtained a cooling age of $\tau_{\rm DA} = 0.74^{+0.09}_{-0.06}$ Gyr, resulting in a total age of $\Gamma'_{\rm DA} = 4.6^{+5.44}_{-2.43}$ Gyr for \wdbdA{}. The derived final WD's mass using this approach was $M_{\rm DA} = 0.61^{+0.05}_{-0.04}~M_\odot$, which is consistent within its uncertainties with previous findings. Furthermore, the results obtained from the SED analysis agree with this estimate, indicating that the WD likely belongs to an older population of stars, formed from a relatively low-mass progenitor, and has undergone a rapid cooling process since its formation (e.g., \citealt{1993A&A...271L..13S, 2000ApJ...528..397I, 2000ApJ...543..216C}). For the non-DA WD case, we obtained similar values for the aforementioned parameters: $\Gamma_{\rm nDA} = 4.36^{+5.54}_{-2.88}$ Gyr, $M_{\rm init, nDA} = 1.3^{+0.6}_{-0.3}~M_\odot$, $\tau_{\rm nDA} = 0.79^{+0.09}_{-0.07}$ Gyr, $\Gamma'_{\rm nDA} = 5.17^{+5.48}_{-2.83}$ Gyr, and $M_{\rm nDA} = 0.60^{+0.05}_{-0.04}~M_\odot$.

\subsection{Ultracool Dwarf's characterisation}

{From the available photometric measurements of \wdbdB{} (Table \ref{tab:photometry}), the best-fit was obtained with the {\sc AMES-Dusty} model \citep{2000ApJ...542..464C, 2001ApJ...556..357A}, yielding an effective temperature of $T_{\rm eff} = 2400\pm50~K$. For the system's distance, its bolometric luminosity was estimated as $L_{\rm bol} = (5.60\pm0.02)\times10^{-4}~L_\odot$, and applying Stefan-Boltzmann's law, \wdbdB{} radius was estimated as $R = 1.22\pm0.05~R_{\rm Jup}$. Within its uncertainties, and since we know precisely the distance of the object, those values are consistent with objects of its kind (e.g., \citealt{1963PThPh..30..460H, 1999ApJ...519..802K, 2001RvMP...73..719B, 2012ARA&A..50...65L}). The UCD's age is unknown, especially due to the degeneracy in the mass-age relationship for sub-stellar objects, therefore, assuming that \wdbdB{} was formed at the same time as its companion, \wdbdA{} ($\Gamma = 4.6^{+5.5}_{-2.4}~{\rm Gyr}$), the mass of the companion was estimated to be $M_{\rm UCD} = 0.094\pm0.006~M_\odot$ ($=98.5\pm6.3~M_{\rm Jup}$), and its surface gravity was consequently estimated at $\log(g)$[cgs] $=5.21\pm0.04$, suggesting an L/M dwarf nature \citep{2005nlds.book.....R, 2013ApJS..208....9P, 2022A&A...660A..38W}. Moreover, we estimated the spectral type of \wdbdB{} as M8 from the empirical stellar spectra library of \cite{2017ApJS..230...16K} within VOSA's framework.

While we designate \wdbdB{} as an ultracool dwarf, we observe that its mass falls $2-\sigma$ within the hydrogen-burning main sequence edge mass (HBBM) of $0.075-0.092~M_{\odot}$ ($\sim78.5-96.3~M_{\rm Jup}$; \citealt{1997A&A...327.1039C, 2001RvMP...73..719B, 2016AdAst2016E..13A}), in similarity to other known low-mass brown dwarfs at similar effective temperatures (e.g., Teide 1; $M = 55\pm15~M_{\rm Jup}$ and $T_{\rm eff} = 2.584\pm150~K$; \citealt{2000ASPC..212...82L}). Spectroscopic observations of \wdbdB{} make itself necessary for its accurate characterisation and classification. 
}

Using Kepler's third law, by considering the masses of the WD and UCD along with their projected separation and face-on orientation, we estimated a minimum orbital period of the \wdbd{} system as $141\pm4$ Kyr. Despite the lengthy orbital period, the system is expected to remain gravitationally bound, with a comparison with Figure 7 of \cite{2012AJ....144..180M} -- stating empirical limits for the stability of a sample of binary systems --, which in turn is based on \cite{2013ApJ...772..129B}, supporting such a conclusion.

\section{On the Membership Likelihood of \wdbd{} to the Sco-Cen Association}\label{sec:membership}

{To assess the likelihood of \wdbd{} pair being a member of the Sco-Cen stellar association, we used the {\sc BANYAN} $\Sigma$ algorithm based on a 6D {\sc XYZUVW} kinematic analysis \citep{2018ApJ...856...23G}, applying both to the LCC stellar association, the Upper Centaurus Lupus (UCL), and Upper Scorpius (US); branches within the Sco-Cen group. This analysis indicates a null probability for the system to belong to the LCC group or any sub-structures within it. This conclusion is also supported by the fact that the inferred ages of the WD component (as shown in Table \ref{tab:wdbdphysics}) do not align with the ages typically associated with the LCC stellar association; previous studies, for instance, by \cite{2018ApJ...868...32G} and \cite{2016MNRAS.461..794P} suggest ages of $10$ Myr and $17$ Myr, respectively.} 

Unlike massive stars that evolve into WDs within a few million years, the formation of a WD from relatively low-mass star progenitors generally requires a significant amount of time for a star to exhaust its nuclear fuel, shed its outer layers, and gradually cool down \citep{2013sse..book.....K}. This process typically takes billions of years, making it improbable for a WD like \wdbdA{} to have an age as young as 17 Myr. This inconsistency between the estimated ages and the established age range adds to evidence that contradicts \wdbd{}'s association with the LCC, supporting that the system is likely to be a field object. 

\begin{figure}
    \centering
    \includegraphics[width = \columnwidth]{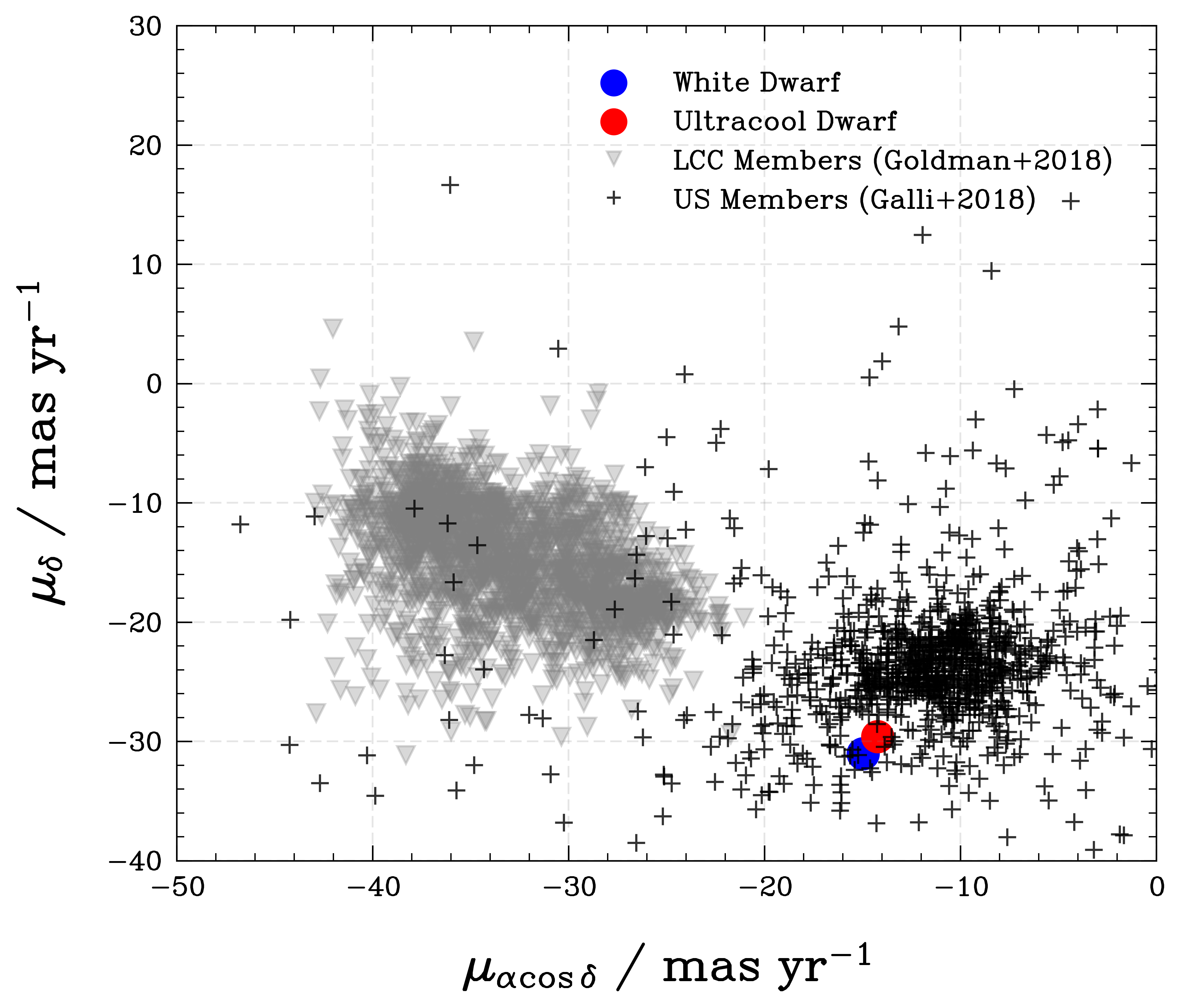}
    \caption{Proper motion distribution of LCC (grey triangles; \citealt{2018ApJ...868...32G}) and US (black crosses; \citealt{2018MNRAS.477L..50G}) members. The components of \wdbd{} are depicted as blue (WD) and red (UCD) circles.}
    \label{fig:GMM}
\end{figure}

\section{Near-IR VVV Light Curves}\label{sec:lcs}

\begin{figure}
    \centering
    \includegraphics[width = \columnwidth]{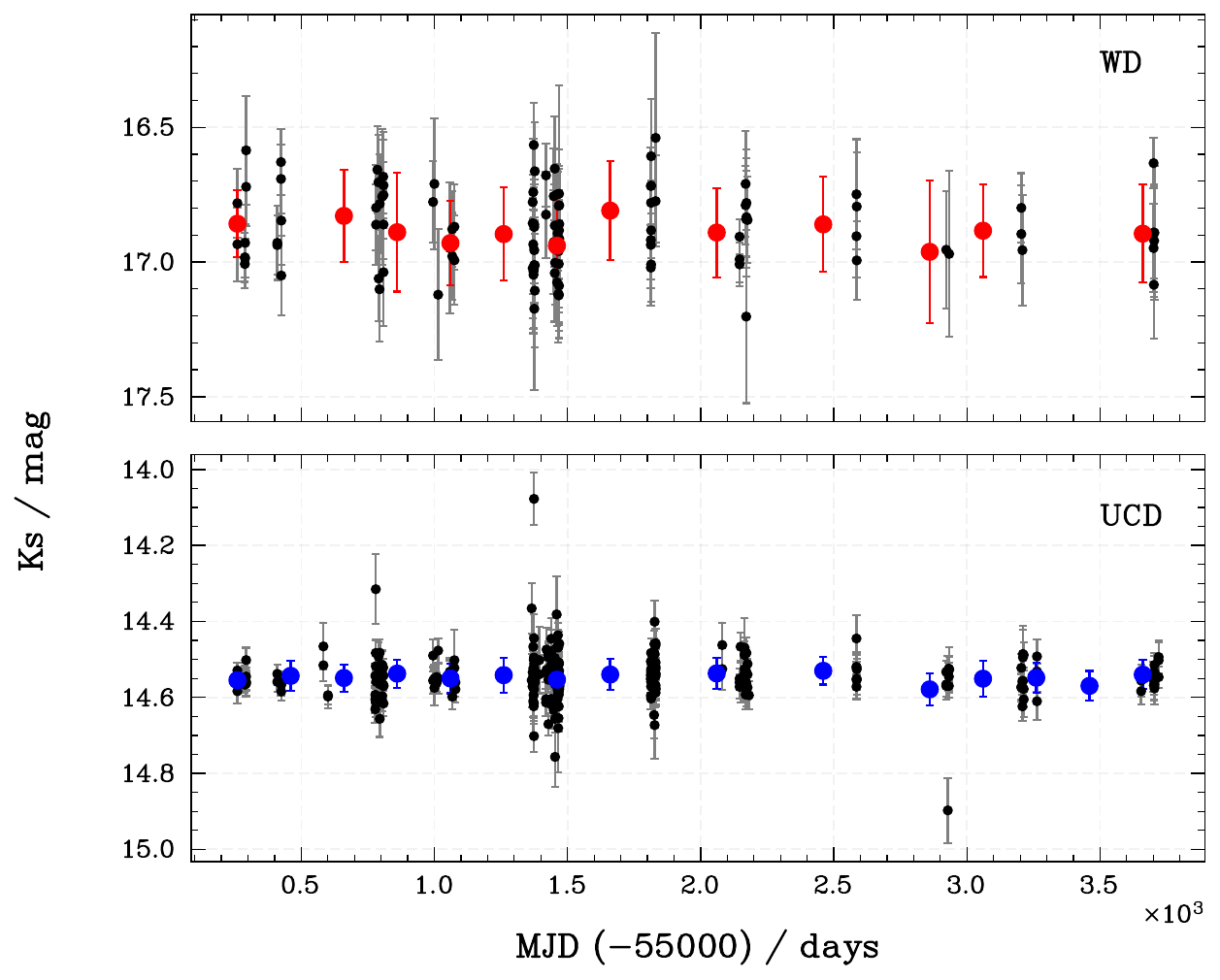}
    \caption{Observed $K_\mathrm{s}-$band near-IR light curves for the WD (top panel) and UCD companions (bottom panel) spanning 10 years. The red dots depict the moving averages for 500 days of steps.}
    \label{fig:VVVLCs}
\end{figure}

Both targets are situated in the overlapping areas of two VVV tiles (d014, centred at $l = 314.40^\circ$, $b = -1.64974^\circ$, and d015, centred at $l = 315.83^\circ$, $b = -1.64972^\circ$), yielding almost twice as many observations concerning single-tile objects. However, in some of these images, the targets are too close to the edge, rendering impossible photometric measurements. Consequently, the final $K_\mathrm{s}$ band light curves consist of 115 photometric data points for the WD (46 in d014 and 69 in d015), and 285 for the UCD companion (104 in d014 and 181 in d015), respectively. In addition, there are 7 measurements in the $Z-$ and $Y-$ bands for both components, plus 11 in the $J-$band and 6/5 measurements in the $H-$band for both the WD/BD (see Table \ref{tab:photometry}). The VVV $K_\mathrm{s}-$filter light curves for the UCD and WD are presented in Figure \ref{fig:VVVLCs}. 

\begin{table}
	\centering
	\caption{VVV near-IR photometry of \wdbd{} pair. The full version of this table is available in the supplementary material, where VVV ID 11846281028924 stands for the WD companion, and VVV ID 11846281000738 stands for the UCD companion.}
	\label{tab:lightcurve}
	\begin{tabular}{llccr}
		\hline
        VVV ID & MJD & Filter & Magnitude & $\sigma$ / mag \\
        \hline 
        11846281028924 & 55260.349798 & K$_\mathrm{s}$ & 16.781 & 0.1271 \\[0.1cm]
        11846281028924 & 55260.350439 & K$_\mathrm{s}$ & 16.933 & 0.1365 \\[0.1cm]
        11846281028924 & 55283.213633 & Y & 16.821 & 0.0148 \\[0.1cm]
        11846281028924 & 55283.215090 & Y & 16.818 & 0.0159 \\[0.1cm]
        11846281028924 & 55283.218185 & Z & 16.826 & 0.0128 \\[0.1cm]
        11846281028924 & 55283.219530 & Z & 16.844 & 0.0138 \\[0.1cm]
        11846281028924 & 55288.357982 & H & 16.854 & 0.0354 \\[0.1cm]
        11846281028924 & 55288.359438 & H & 16.865 & 0.0296 \\[0.1cm]
        11846281028924 & 55288.362746 & K$_\mathrm{s}$ & 17.007 & 0.0886 \\[0.1cm]
        $\dots$ & $\dots$ & $\dots$ & $\dots$ \\[0.1cm]
        \hline
        11846281000738 & 55260.349798 & K$_\mathrm{s}$ & 14.528 & 0.0191 \\[0.1cm]
        11846281000738 & 55260.350439 & K$_\mathrm{s}$ & 14.583 & 0.0311 \\[0.1cm]
        11846281000738 & 55283.213633 & Y & 16.043 & 0.0081 \\[0.1cm]
        11846281000738 & 55283.215090 & Y & 16.052 & 0.0079 \\[0.1cm]
        11846281000738 & 55283.218185 & Z & 16.955 & 0.0192 \\[0.1cm]
        11846281000738 & 55283.219530 & Z & 16.936 & 0.0169 \\[0.1cm]
        11846281000738 & 55288.357982 & H & 14.877 & 0.0126 \\[0.1cm]
        11846281000738 & 55288.359438 & H & 14.872 & 0.0083 \\[0.1cm]
        11846281000738 & 55288.362746 & K$_\mathrm{s}$ & 14.551 & 0.0357 \\[0.1cm]
        $\dots$ & $\dots$ & $\dots$ & $\dots$ \\[0.1cm]
        \hline 
	\end{tabular}
\end{table}

We examined periodic patterns in the VVV $K_\mathrm{s}-$band light curves of both the companions of the \wdbd{} pair within a range of $0.1$ days up to half of the full baseline with the Generalised Lomb-Scargle (GLS; \citealt{2009A&A...496..577Z}), Box-fitting Least Squares (BLS; \citealt{2002A&A...391..369K}), Phase Dispersion Minimisation (PDM; \citealt{1978ApJ...224..953S}), and Multi-Harmonic Analysis of Variance (MHAoV; \citealt{1996ApJ...460L.107S}) algorithms through the {\sc astrobase} software \citep{2021zndo...4445344B}. 

\wdbdA{} displays variability with a median absolute variation of 0.08 mag and root mean square of 0.13 mag at a mean magnitude of 16.88 mag. The period search analysis reveals a prominent modulation at 0.112 days ($\sim$161.2 minutes) with the GLS method, which could be associated with periodic patterns caused by surface magnetic variations on the WD that become visible as it rotates, e.g., \cite{2015ApJ...814L..31K, 2021ApJ...923L...6K, 2022AJ....164..131W}. The other period-search algorithms yielded similar periods of 0.126 and 0.109 days for the PDM and BLS algorithms, while for the MHAoV a prominent 2.076-day period was found. Given the relatively noisy nature of the resulting periodograms (see Appendix \ref{app:periodograms}), further investigation is needed to unveil the nature of these modulations. For \wdbdB{}, slight variability was noticed in the $K_\mathrm{s}-$band, suggesting two potential modulation periods: $1.288$ days with the LS method and $1.729$ days with the PDM method. None of these appears significant given the noisy nature of the resulting periodogram (see Appendix \ref{app:periodograms}) and no clear sine-like variations at these periods. The UCD phase diagram recovered with MHAoV at a period of 7.73 days does not exhibit any discernible temporal variation, while applying the BLS algorithm, a modulation period of $P_{\rm \tiny BLS, UCD} = 0.111$ days was recovered, {albeit also with low-significance due to inherent noise in the periodogram. These observed modulations might be influenced by various factors, such as errors in measurements or issues with the measuring instruments, resulting in stochastic fluctuations in the data. We also note the possibility of false periods in the periodogram, potentially arising from systematic effects introduced during the data windowing process, which could involve hidden periods or non-stationary behaviour. Further investigation is also needed to unveil the nature of these modulations.}

\section{Discussion and Conclusions}\label{sec:conclusions}

{We present the characterisation of \wdbd{}, a wide binary system consisting of a WD and a UCD first discovered with VVV/VVVX data analysis during searches for companions to UCDs, BDs and FFPs towards the Sco-Cen/LCC stellar association \cite{2021MNRAS.506.2269E} and independently by \cite{2022cosp...44..591M}. The SED analysis of this system has been conducted from data of the \emph{Gaia} DR3, VVV and DECaPS survey within VOSA, to derive the physical, atmospheric, and dynamical parameters of both components. Space velocities demonstrate that the system belongs to the thin disc and is gravitationally bounded from the considerably small chance alignment probability. The DA-type WD has an effective temperature of $9500\pm125~K$, a surface gravity (cgs) of $8.00\pm0.12$, a mass of $0.62\pm0.18~M_\odot$, and a radius of $0.01309\pm0.0003~R_\odot$. The M8-type UCD, on the other hand, have an effective temperature of $2400\pm50~K$, surface gravity (cgs) of $5.20\pm0.04$, mass of $98.5\pm6.2~M_{\rm Jup}$, and radius of $1.22\pm0.05~R_{\rm Jup}$. These parameters are consistent with the expected characteristics of objects of their respective kinds. Spectroscopic confirmation for both members of the \wdbd{} pair is encouraged and necessary, especially to accurately determine the spectral type of \wdbdB{} beyond its photometric classification, and to assess its precise age.}

\wdbd{} stands out as a benchmark system for comprehending the evolution of WDs and their low-mass companion objects given several notable features, along with the scarcity of such reported binary types. Considering the WD's nominal mass to present $0.62~M_\odot$, the mass ratio ($q \equiv M_{\rm UCD}/M_{\rm WD}$) of \wdbd{} can be estimated as $0.15\pm0.04$, which suggests a binary-like formation scenario for this system, i.e., the two objects form independently, and their masses might not be closely related \citep{2014prpl.conf..619C, 2020AJ....159...63B}. {The co-natal status of the pair, as well as for other WD+UCD/MS and low-mass double binaries -- both wide and close -- contingent upon their gravitational binding and co-mobility, remains a subject of ongoing debate and will be investigated in a future study. \cite{2021ApJ...921..118N}, for instance, demonstrated that $\sim73\%$ of the far ($2\times10^{5}-10^{7}$ au; 31 pairs) co-moving MS+MS binaries are co-natal, which is in-line with theoretical predictions from \cite{2019ApJ...884..173K}.} The total age of \wdbdA{} was estimated at $4.6^{+5.5}_{-2.4}$ Gyr (cooling age of $0.73^{+0.09}_{0.06}$ Gyr), and given the long orbital period of the system ($141.8\pm4.5$ Kyr) plus loose separation ($\sim 1236.73~{\rm au}$), it indicates a detached coexistence, with minimal perturbations or interchanges, inviting further scrutiny into the intricate interplay of forces that have governs the evolution of WD+UCD/BD/MS' evolution.

\section*{Acknowledgements} 

We would like to thank the referee for their important comments on the manuscript, which have certainly helped to improve the quality of our work. We acknowledge the use of data from the ESO Public Survey program IDs 179.B-2002 and 198.B-2004 taken with the VISTA telescope and data products from the CASU. T.F. acknowledges support from CAPES/Brazil (88887.638119/2021-00) and Yale GSAS. R.K.S. acknowledges support from CNPq/Brazil through projects 308298/2022-5, 350104/2022-0 and 421034/2023-8. D.M. gratefully acknowledges support from the ANID BASAL projects ACE210002 and FB210003, from Fondecyt Project No. 1220724, and from CNPq Brasil Project 350104/2022-0. C.C. acknowledges support by ANID BASAL project FB210003 and ANID, -- Millennium Science Initiative Program -- NCN19$\_{171}$. J.A.-G. acknowledges support from Fondecyt Regular 1201490, and ANID – Millennium Science Initiative Program – ICN12\_009 awarded to the Millennium Institute of Astrophysics MAS. This publication makes use of VOSA, developed under the Spanish Virtual Observatory through grants PID2020-112949GB-I00, and no. 776403 (EXOPLANETS-A from the European Union's Horizon 2020 Research and Innovation Programme).

\section*{Data Availability}

The data used in this project are public and can be accessed/retrieved through the VISTA Science Archive (\url{http://horus.roe.ac.uk/vsa/index.html}), and the Virtual Observatory SED Analyser (\url{http://svo2.cab.inta-csic.es/theory/vosa/index.php}).

\appendix

\section{Periodograms and Spectral Window Function for \wdbd{} System}\label{app:periodograms}

\begin{figure*}
    \centering
    \includegraphics[width = \linewidth]{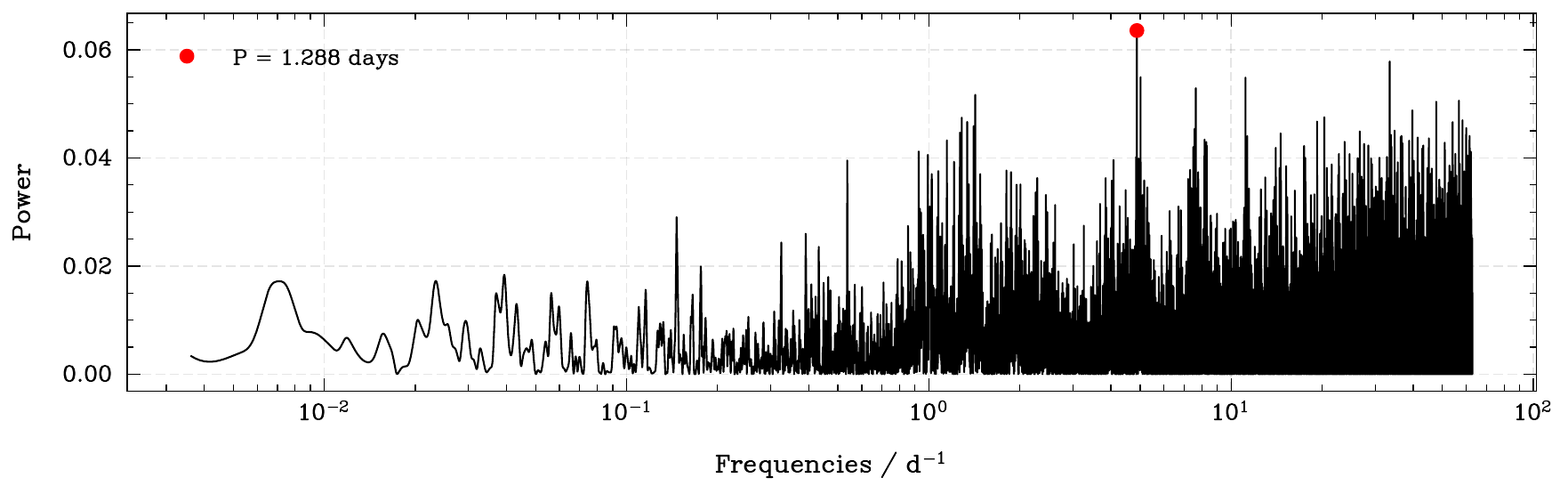}
    \includegraphics[width = \linewidth]{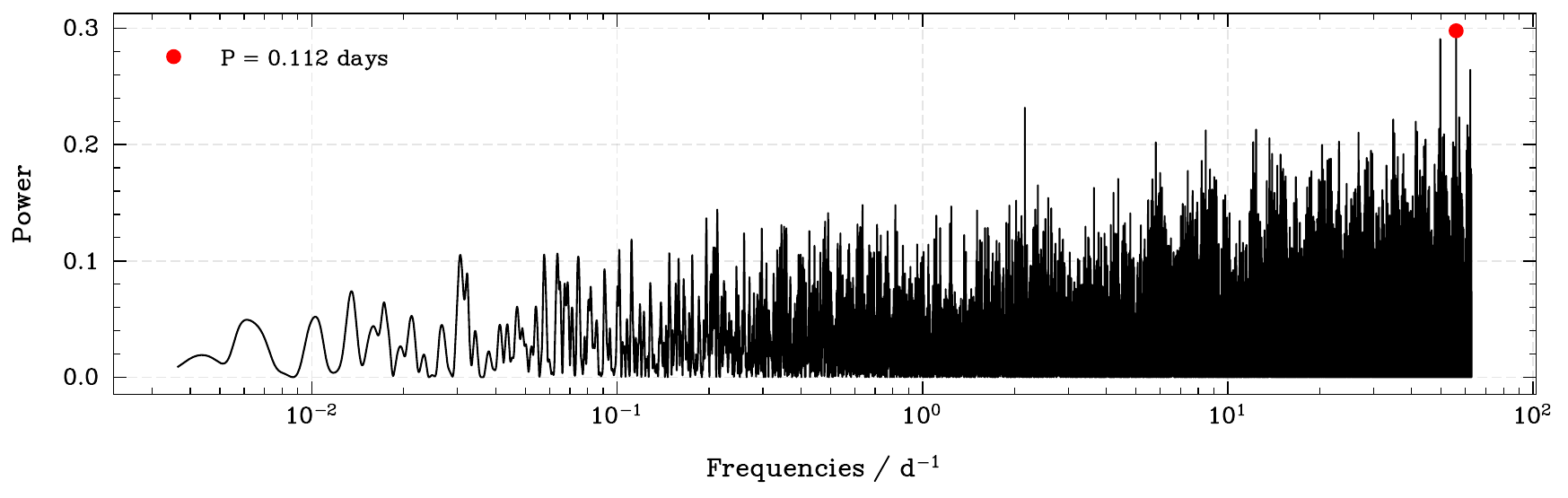}\\
    \includegraphics[width = \linewidth]{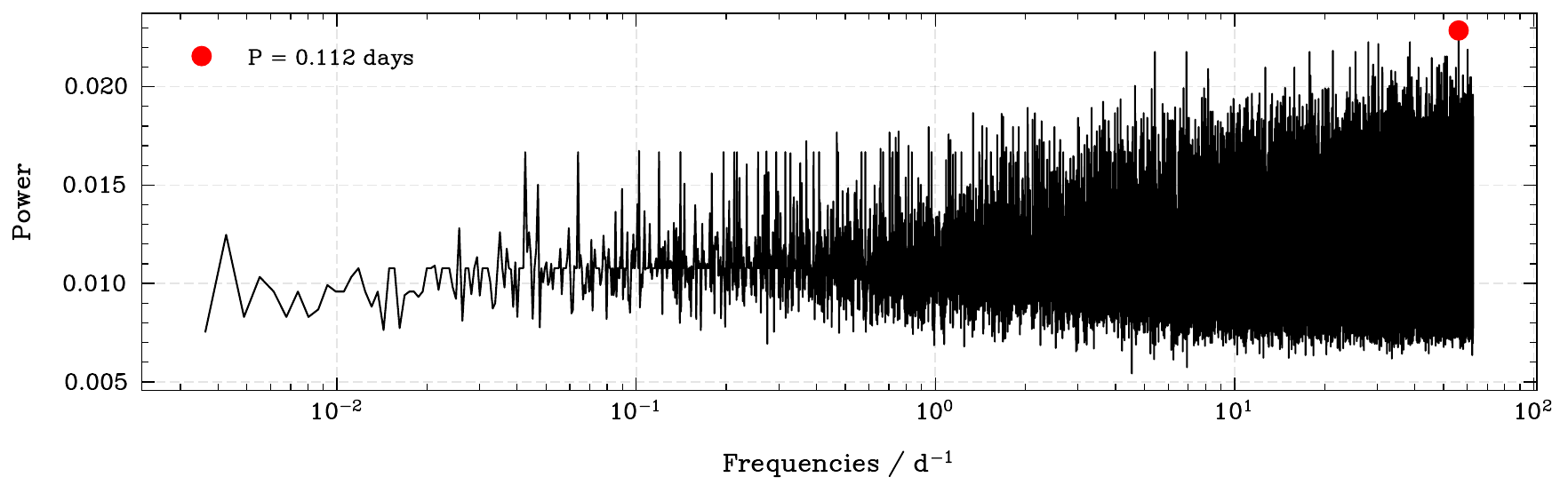}
    \includegraphics[width = \linewidth]{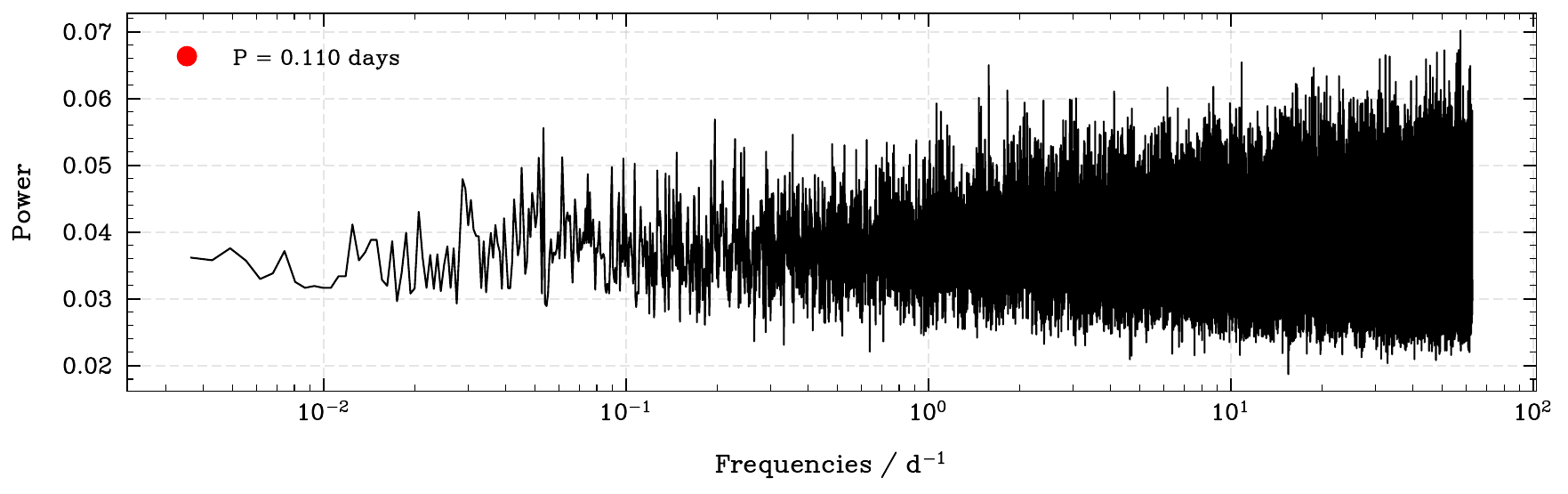}
    \caption{Lomb-Scargle (first and second panels) and Box Least-Squares (third and fourth panels) periodograms of the VVV $K_\mathrm{s}$ time-series of the \wdbdB{} and \wdbdA{}, respectively.}
    \label{fig:KBLD+LS_periodograms}
\end{figure*}

\begin{figure*}
    \centering
    \includegraphics[width = \linewidth]{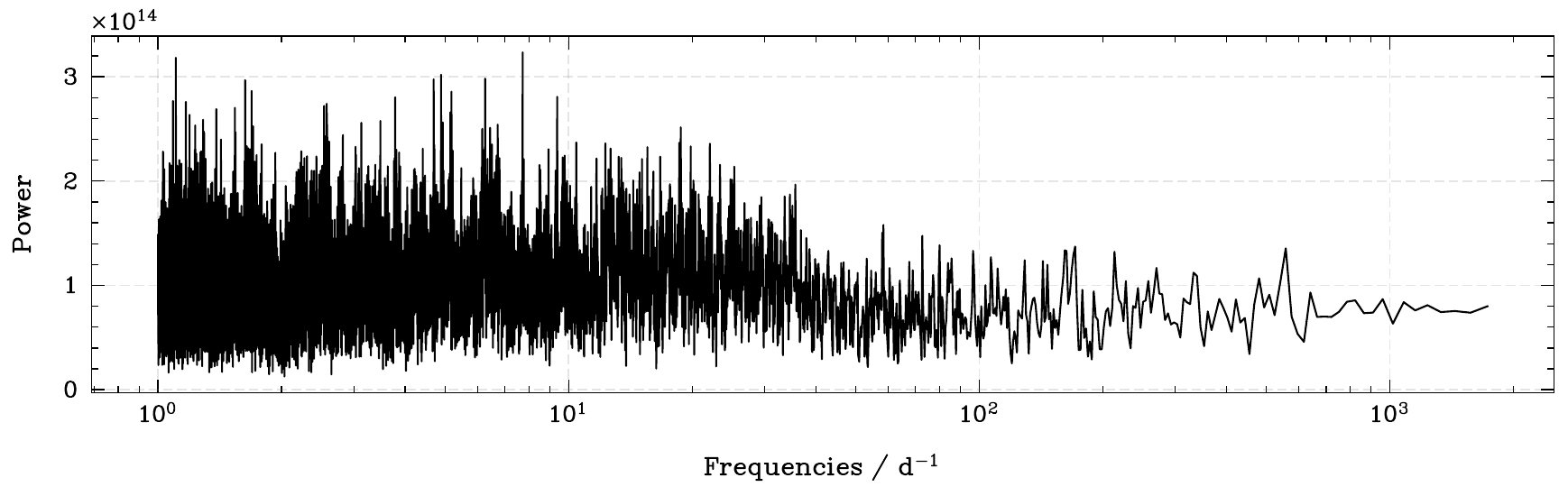}
    \includegraphics[width = \linewidth]{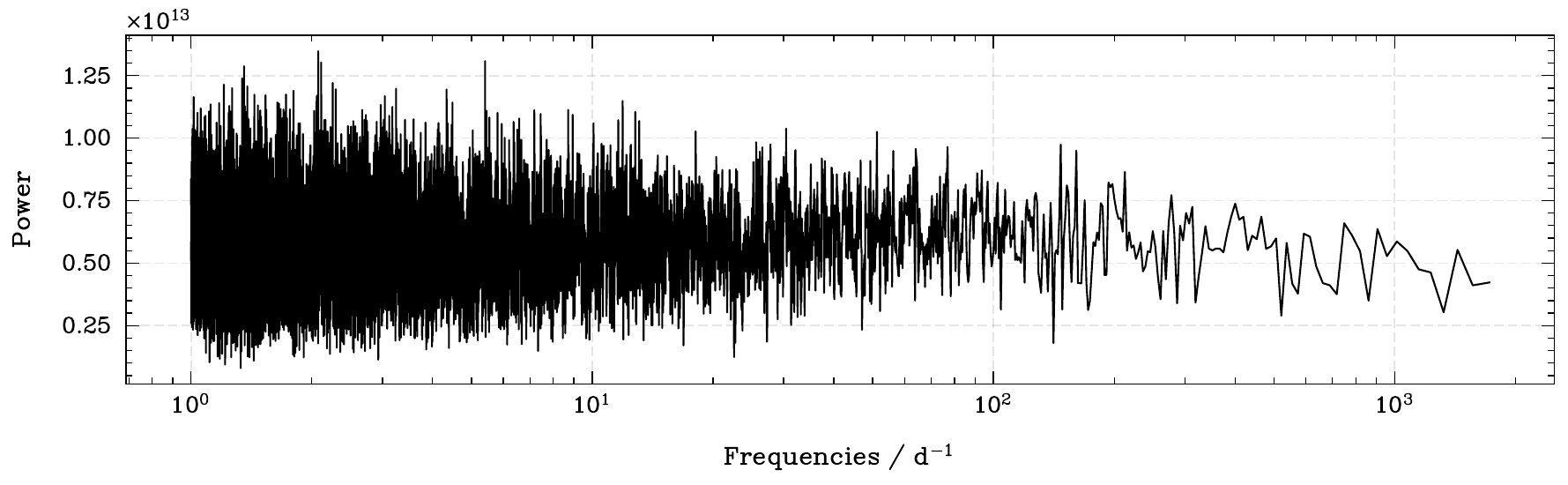}\\
    \includegraphics[width = \linewidth]{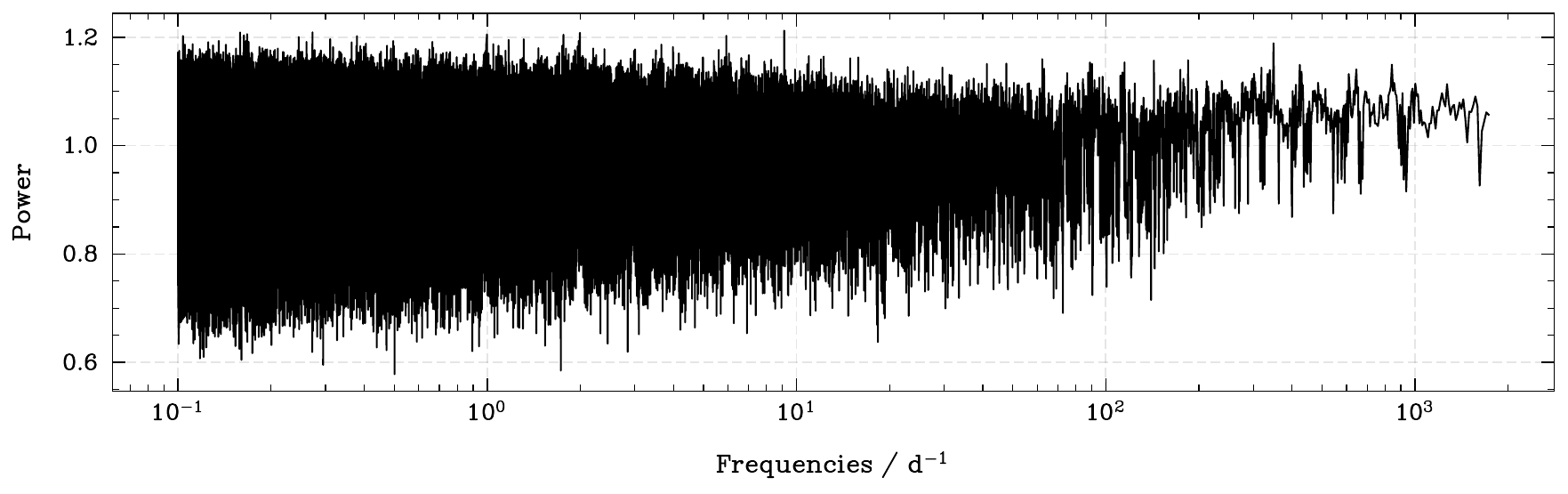}
    \includegraphics[width = \linewidth]{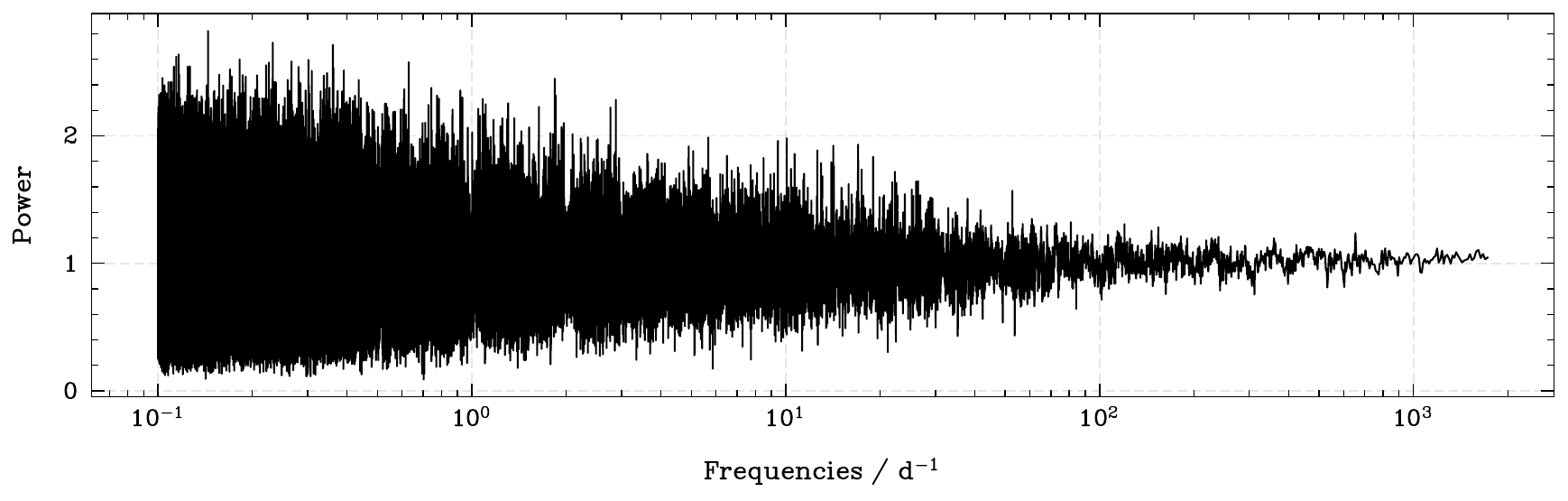}
    \caption{Multi-Harmonic Analysis of Variance (first and second panels) and Phase Dispersion Minimisation (third and fourth panels) periodograms of the VVV $K_\mathrm{s}$ time-series of the \wdbdB{} and \wdbdA{}, respectively.}
    \label{fig:PDM+AOV_periodograms}
\end{figure*}

\begin{figure*}[h]
    \centering
    \includegraphics[width = \linewidth]{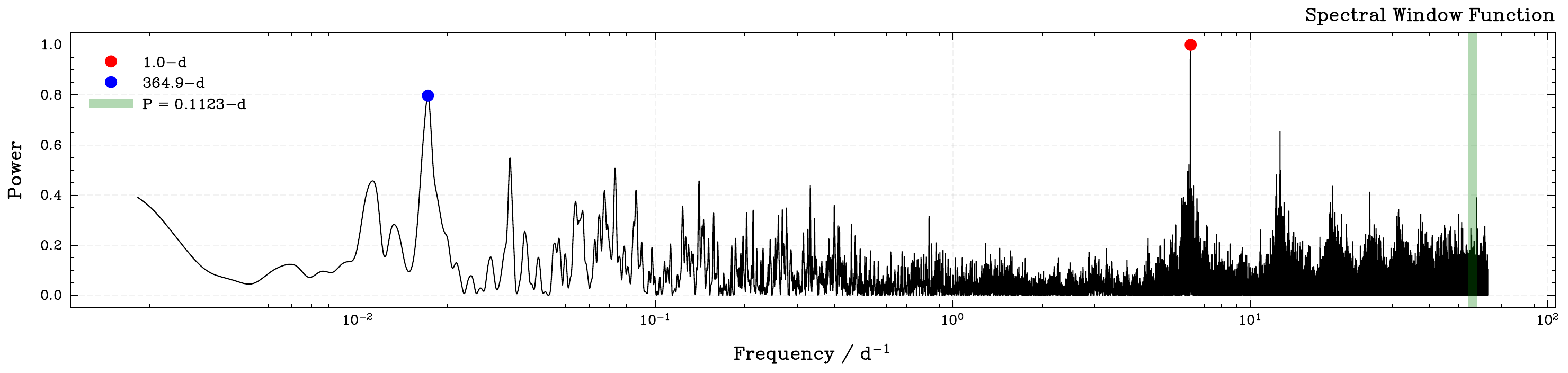}\\
    \includegraphics[width = \linewidth]{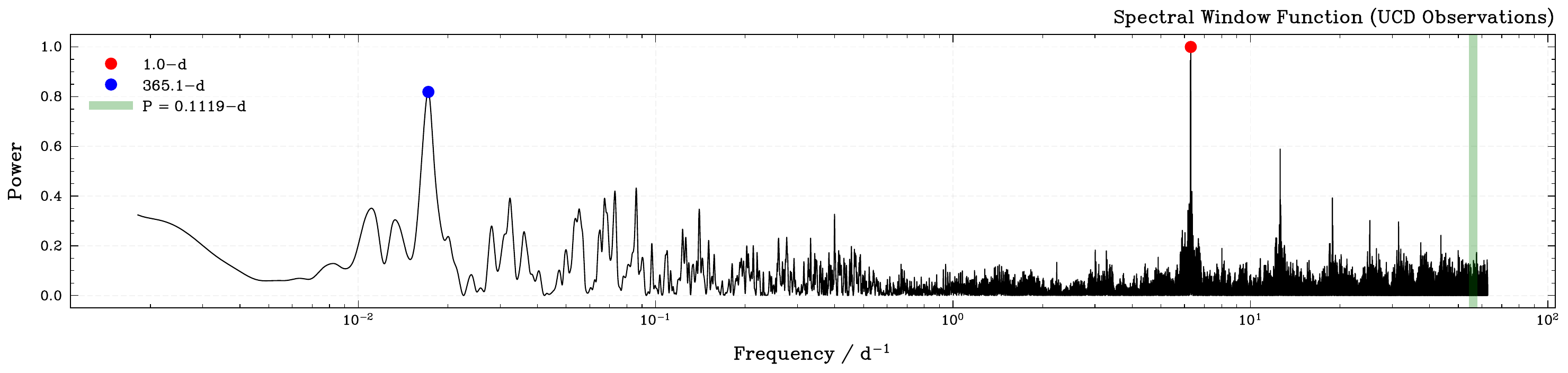}
    \caption{Spectral Window Function for \wdbdB{} and \wdbdA{} observations, respectively. The green vertical line marks the prominent modulation period found with the LS and BLS for both the WD and the UCD components.}
    \label{fig:SWF}
\end{figure*}

\bsp \label{lastpage} \end{document}